\documentclass[sigconf]{acmart}

\AtBeginDocument{%
  }

\usepackage{amsmath}

\usepackage{amssymb}
\usepackage{graphicx}
\usepackage{booktabs}
\usepackage[table]{xcolor}
\usepackage{subcaption}
\usepackage{enumitem}
\usepackage{listings}
\usepackage{siunitx}
\begin{document}
\acmYear{2026}\copyrightyear{2026}
\setcopyright{cc}
\setcctype[4.0]{by}
\acmConference[CAIS '26]{Proceedings of the 1st ACM Conference on Agentic and AI Systems}{May 26, 2026}{San Jose, CA, USA}
\acmBooktitle{Proceedings of the 1st ACM Conference on Agentic and AI Systems (CAIS '26), May 26, 2026, San Jose, CA, USA}
\acmDOI{10.1145/3786335.3813188}
\acmISBN{979-8-4007-2415-2/26/05}

\title{HearthNet: Edge Multi-Agent Orchestration for Smart Homes}

\author{Zhonghao Zhan}
\affiliation{%
  \institution{Imperial College London}
  \city{London}
  \country{UK}}
  \email{zzhan@ic.ac.uk}

\author{Krinos Li}
\affiliation{%
  \institution{Imperial College London}
  \city{London}
  \country{UK}}
  \email{k.li23@imperial.ac.uk}

\author{Yefan Zhang}
\affiliation{%
  \institution{Independent Researcher}
  \city{Seattle}
  \country{USA}}
  \email{zhangyefan752@gmail.com}

\author{Hamed Haddadi}
\affiliation{%
  \institution{Imperial College London}
  \city{London}
  \country{UK}}
  \email{h.haddadi@imperial.ac.uk}

\begin{abstract}
Smart-home users increasingly want to control their homes in natural language rather than assemble rules, dashboards, and API integrations by hand. At the same time, real deployments are brittle: devices fail, integrations break, and recoveries often require manual intervention. Existing agent toolkits are effective for session-scoped delegation, but smart-home control operates under a different scenario: it is persistent, event-driven, failure-prone, and tied to physical devices with no shared context window. We present \textsc{HearthNet}, an edge multi-agent orchestration system for smart homes\footnote{
Demo video: \url{https://www.youtube.com/watch?v=p3ZKDsKifRk}; Interactive Demo: \url{https://hearthnet.vercel.app/}}
. \textsc{HearthNet} deploys a small set of persistent, role-specialized LLM agents at the home hub, where they coordinate through MQTT, Git-backed shared state, and root-issued actuation leases to govern heterogeneous devices through thin adapters. This design externalizes context, preserves execution history, and separates planning, verification, authorization, and actuation across explicit boundaries. Our current prototype runs on commodity edge hardware and Android devices; it keeps orchestration, state management, and device control on-premise while using hosted LLM APIs for inference. We demonstrate the system through three live scenarios: intent-driven multi-agent coordination from ambiguous natural language, conflict resolution with timeline-based tracing, and rejection of stale or unauthorized commands before device actuation.

\end{abstract}



\maketitle

\section{Introduction}

AI agents are moving beyond cloud sandboxes into persistent physical environments such as home servers, IoT hubs, and personal devices. This shift exposes a mismatch between current orchestration toolkits and the realities of edge deployment.

Dominant multi-agent frameworks, including Anthropic's agent teams, Microsoft's AutoGen, and LangChain's LangGraph, expose state-serialization APIs but leave durable persistence and cross-machine coordination to the application. Anthropic notes that subagents remain tied to a session context window and that cross-session coordination requires a different pattern~\cite{anthropic-agents}. Microsoft notes that AutoGen leaves persistence and recovery to the application~\cite{ms-agentframework-migration}. LangChain similarly warns that once agents become distinct graph nodes, context transfer becomes an explicit engineering problem~\cite{langchain-multiagent}.

Smart-home control operates under a different regime. Coordination is long-running, and physically consequential: devices crash, networks partition, batteries die, and stale or conflicting commands can create safety, privacy, or usability failures. These properties demand explicit shared state, recovery semantics, and permission boundaries that session-scoped subagents do not naturally provide.

When orchestration spans heterogeneous devices, multiple privilege domains, and recovery from failures, a small set of long-lived specialist agents with explicit shared state is easier to audit, recover, and resume than a monolithic agent or transient subagents.  

We present \textsc{HearthNet}, an edge multi-agent orchestration system for smart homes. \textsc{HearthNet} deploys a small set of persistent, role-specialized LLM agents on commodity edge hardware, coordinated through MQTT, Git-backed shared state, and root-mediated authorization. The prototype keeps orchestration, state management, and device control on-premise while using hosted LLM APIs for inference. We make four contributions:

\begin{itemize}[nosep,leftmargin=*]
  \item A \textbf{Git-backed freshness mechanism} (base commit
    verification) that enables agents to detect and reject stale or
    conflicting commands before device actuation.
  \item \textbf{Policy-bound actuation leases} in which the root agent
    issues short-lived, machine-checkable authorizations binding each
    state-changing command to an identity, device scope,
    parameter envelope, and the state version under which it was
    approved.
  \item A \textbf{librarian-as-observer} pattern in which a dedicated
    non-participant agent mirrors coordination traffic into a versioned
    repository, providing crash recovery, attribution, and full
    auditability without participating in the control path.
  \item A \textbf{deployed edge prototype} running on three classes of
    commodity hardware (ARM64 Mac mini, x64 NUCs, Android phone),
    demonstrated through three live scenarios with real devices.
\end{itemize}

\section{System Architecture}
\label{sec:arch}

\textsc{HearthNet} uses a hierarchical role architecture with four tiers
(Figure~\ref{fig:arch}): a \emph{root agent}, \emph{manager agents},
\emph{device adapters}, and a cross-cutting \emph{librarian agent}. All
inter-agent coordination flows over a shared MQTT broker, while durable
shared state is maintained in a Git repository managed by the librarian.

A typical request begins at the root agent, which receives a user
command, decomposes it into domain-specific tasks, and dispatches them
to one or more managers over MQTT. Managers resolve the request against
the current Git-backed state, prepare concrete device actions, and
return execution requests or conflict reports. Before any state-changing
operation reaches a device adapter, the root agent verifies freshness,
checks policy, and issues a short-lived actuation lease. The librarian
mirrors the resulting coordination traffic and records externally
relevant events into the shared repository for audit and recovery \footnote{All code, configuration, agent templates, and replay scripts are
available at \url{https://github.com/zhonghaozhan/hearthnet}.}.

\subsection{Roles and Communication}

\begin{figure}[t]
  \centering
  \includegraphics[width=\columnwidth]{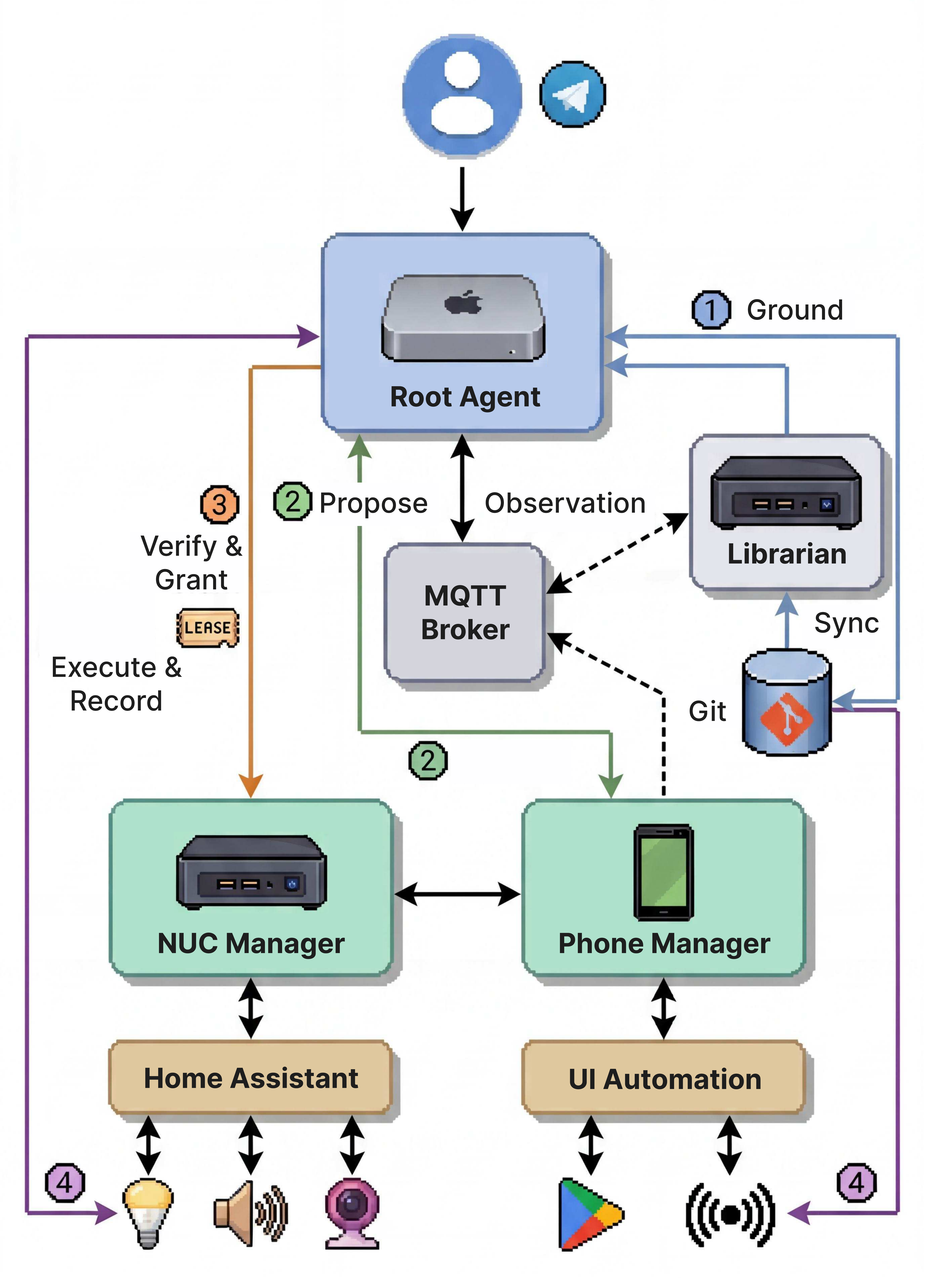}
  \caption{\textsc{HearthNet} architecture. Solid arrows indicate the command path; the librarian mirrors coordination traffic and records externally relevant events to Git.}
  \Description{Four-tier diagram showing a root agent (Rupert) on a Mac mini at the top, dispatching tasks over MQTT to two manager agents (Jeeves on an Intel NUC, Darcy on an Android phone) which control device adapters; a separate librarian agent (Dewey) on another NUC mirrors all MQTT traffic into a Git repository for audit and recovery.} 
  \label{fig:arch}
\end{figure}

Each major role runs as an independent edge-resident runtime, on its own device, hosted using OpenClaw~\cite{openclaw} as the agent
runtime platform. Agents coordinate over a central Mosquitto MQTT
broker using point-to-point inboxes, a broadcast channel, and a mirrored
audit stream. Messages are JSON-encoded and include an optional
\texttt{base\_commit} hash referencing the Git state the sender believes
to be current. State-changing execution requests additionally carry a
lease id referencing a root-issued authorization record (Figure~\ref{fig:arch}).

\paragraph{Root Agent.}
\textbf{"Rupert"} on a Mac Mini serves as the system's top-level
orchestrator. It receives high-level user commands via Telegram,
decomposes them into domain-specific tasks, dispatches them to managers,
arbitrates conflicts, and is the only component allowed to issue
actuation leases. Rupert holds the broadest system context but
delegates device-specific resolution to managers.

\paragraph{Manager Agents.}
Managers each govern a domain of devices and report to Rupert:
\begin{itemize}[nosep,leftmargin=*]
  \item Home-Assistant / NUC Manager (\textbf{"Jeeves"}). Runs on an x64 Intel
    NUC and manages devices exposed through Home
    Assistant~\cite{home-assistant}, including lights, switches, climate,
    and media endpoints.
  \item Mobile-App / Phone Manager (\textbf{"Darcy"}). Runs on a Pixel 4
    (Android 14) phone and controls proprietary IoT applications that expose no external API, using UI automation through ADB and accessibility
    services. Darcy also accesses on-device sensors (camera, ambient
    light, gyroscope) that are physically unavailable to hub-resident
    agents, a key reason a single centralized agent cannot replace the
    distributed architecture.
\end{itemize}
Managers can propose device actions within their domain, but they do
not hold standing authority to actuate devices; every state-changing
command requires a fresh lease from Rupert.

\paragraph{Librarian Agent.}
\textbf{"Dewey"} runs on a separate Intel NUC and maintains a Git repository as the canonical shared record of system coordination state. It subscribes to mirrored MQTT traffic and records externally relevant events with structured metadata (timestamp, sender, outcome). The events include task dispatches, device-state transitions, conflict resolutions, and recovery outcomes. Dewey
does not issue commands and holds no device credentials; its role is
purely observational. The repository provides three properties:
(1)~\emph{versioned shared state} that agents can inspect and recover
after restart; (2)~\emph{attribution and history} preserved in a
diffable timeline; and (3)~\emph{stale-state detection} via
\texttt{base\_commit} comparison against the current HEAD.

\paragraph{Device Adapters.}
Devices are \emph{not} agents. Lights, speakers, cameras, and sensors
are thin, deterministic endpoints controlled by managers through
protocol-specific adapters. This keeps the agent count small and avoids
an ``agent per device'' design.

\subsection{Authorization Model}
\label{sec:auth}

\textsc{HearthNet} uses root-mediated authorization to keep policy
separate from device logic. A versioned policy file stored in Dewey's
repository maps each manager role to a set of allowed device classes,
operation types, and optional parameter bounds. Before any state change,
Rupert evaluates the requested action against this policy and, if
permitted, issues a short-lived \emph{actuation lease}.

Each lease contains: (1) the grantee manager identity; (2) the target
device or device class; (3) the permitted operation and parameter
envelope; (4) the current \texttt{base\_commit}; (5) the policy commit
under which the decision was made; (6) an expiration time; and
(7) a human-readable justification for audit. Device adapters are
lease-aware: they reject commands whose lease is missing, expired, out
of scope, or bound to an outdated commit. This turns authorization into
a checkable boundary rather than an informal prompt.

\subsection{Execution Protocol}
\label{sec:decision-loop}

Each request follows a concrete four-stage protocol:

\begin{enumerate}[nosep,leftmargin=*]
  \item \textbf{Ground.} The receiving agent loads the current device
    shadow state~\cite{aws-device-shadow}, relevant policy snapshot, and
    repository HEAD from Dewey's Git repository. This ensures reasoning
    is anchored to current system state rather than session-local memory.
  \item \textbf{Propose.} Rupert decomposes user intent into subtasks,
    while managers translate subtasks into concrete device actions. At
    this stage, managers propose commands but do not actuate.
  \item \textbf{Verify and grant.} Rupert checks whether each proposed
    action is fresh with respect to the current HEAD, consistent with
    current intent and conflict rules, and authorized by the current
    policy. For approved actions, Rupert issues an actuation lease.
  \item \textbf{Execute and record.} The manager invokes the adapter
    using the lease, returns the outcome over MQTT, and Dewey records
    the proposal, authorization decision, and result in Git.
\end{enumerate}

Agents are persistent processes, but each incoming message triggers a fresh inference call~\cite{anthropic-long-running} grounded in the current Git state, so a restarted agent behaves identically to a continuously-running one.  

\subsection{Failure Handling}
\label{sec:failure}

Agents publish periodic heartbeats; the MQTT broker's Last Will and
Testament (LWT) feature automatically announces unexpected
disconnections.  The root agent monitors these channels; if a manager's
heartbeat is missing for two consecutive intervals (default: 60\,s),
the root marks it as unresponsive.

Task recovery:
When a manager fails mid-routine, the root agent examines Dewey's Git
log to classify each dispatched subtask as \emph{confirmed},
\emph{in-flight}, or \emph{blocked}. Confirmed tasks are closed.
In-flight tasks are reissued only if another live manager advertises the
required capability and Rupert can issue a new lease under the current
state and policy snapshot. Otherwise the task enters a retry queue with
exponential backoff or is surfaced for operator review. Recovery thus
depends on explicit capabilities and current authorization state, not on
reconstructing hidden conversational context.

State recovery after restart:
When a crashed agent restarts, it pulls the current state from Dewey's
Git repository and includes the current HEAD as \texttt{base\_commit}
in its first message. If a recovering agent sends a command referencing
a stale commit or reuses an expired or invalid lease, the root agent
rejects it and the agent must re-synchronize and request a new
authorization. This will be exactly what Scene~3 of the demonstration
shows. Agent state is never authoritative; the Git repository is the
single source of truth.

\section{Demonstration \& Evidence}
\label{sec:demo}

The live demonstration consists of three scenarios executed end-to-end
in approximately four minutes on the deployed prototype with no
simulation or mocking. Attendees interact with the system through a
Telegram group chat and observe physical device responses in real time.
The three scenes are chosen to exercise the system's core capabilities
in sequence: intent interpretation with multi-agent decomposition
(Scene~1), conflict arbitration grounded in durable state (Scene~2),
and safety enforcement through freshness and authorization verification
(Scene~3).

\subsection{Scene 1: Coordinated Actuation}

An attendee sends an ambiguous, intent-level command via Telegram:
\emph{``I'm working from home today.''} No predefined rule exists
for this request. Rupert interprets the intent, reasons about what
``work from home'' implies for each device domain, and dispatches
domain-specific subtasks: Jeeves sets lights to a club-style tone ($<$60\% brightness) and adjusts the speaker to a low background level ($<$30\% volume); Darcy launches a smart
TV control app on the phone via UI automation and turns off the TV.
Each manager independently translates its subtask into device-specific
actions using its own domain knowledge. Rupert, as the root agent, then
issues a short-lived lease for each approved actuation, binding the
command to the current commit, the manager identity, and the allowed
parameter range before execution proceeds. Dewey commits the full
coordination chain, including lease decisions, to Git.

\textbf{What the audience sees:} Lights shift to a club-style tone, and
the phone launches the control app to turn off the TV. The Telegram chat
shows Rupert's decomposition reasoning and each manager's
confirmation. A \texttt{git log} displays the complete chain with
timestamps, attribution, and issued leases, highlighting that achieving
the same behavior with conventional automation would require substantial
handcrafted cross-device rule engineering.

\subsection{Scene 2: Conflict Resolution with Timeline Tracing}

While the work-from-home configuration from Scene~1 is still active, a
scheduled ``evening wind-down'' routine fires automatically: Jeeves
proposes dimming lights to 20\% and switching to a warm tone.  This
conflicts with the bright lighting that the user explicitly
requested minutes earlier. Rupert detects the conflict (same device,
incompatible states), queries the Git timeline maintained by Dewey, and
arbitrates: the work-from-home mode was an explicit user request; the
wind-down trigger is a time-based schedule that should not override
active user intent.  Rupert maintains the current lighting state and
Dewey commits the resolution with reasoning.

\textbf{What the audience sees:} The scheduled command appears in the
Telegram group alongside the conflict alert.  Rupert explains its
decision with timeline evidence, and the lights remain unchanged.  A
\texttt{git diff} shows the full decision trail is recoverable.

\subsection{Scene 3: Freshness and Authorization Verification}

This scene demonstrates the combined safety gate formed by
\texttt{base\textunderscore commit} freshness checks and actuation leases.
A manager agent is restarted mid-demo to simulate a crash recovery.
Upon restart, the agent re-synchronizes from Dewey's Git repository but
attempts to replay a pre-crash command using an outdated
\texttt{base\_commit} and the lease that had been issued for the earlier
state snapshot. Rupert compares the referenced hash against the current
repository HEAD, determines the relevant state has advanced (due to
Scene~1 and~2 activity), and rejects the command before actuation. Dewey
records the rejection with the invalidating commit and lease identifier.
The recovering agent must pull the current state and request a fresh
lease before retrying.

\textbf{What the audience sees:} An agent goes offline and comes back.
Its first command is visibly rejected in the Telegram group, with the
rejection reason referencing the specific commit and lease that
invalidated the stale state. A \texttt{git log} confirms the state
updates from Scenes~1--2 that the crashed agent missed, making the
safety mechanism concrete and auditable.

\subsection{Evidence}

Agent frameworks (e.g., AutoGen, LangGraph) expose state-serialization APIs but leave durable shared state, freshness checks, and per-action authorization to the application. An agent has no built-in way to verify that another has already actuated a device. Conventional rule-based automation persists device state but lacks authorization boundaries between routines and cannot
arbitrate conflicts using semantic context. \textsc{HearthNet} combines
persistent shared state (Git), per-action authorization
(leases), and intent-aware arbitration (timeline queries),
providing crash recovery and full auditability without
sacrificing the flexibility of LLM-based coordination.

Table~\ref{tab:evidence} summarizes \textsc{HearthNet} orchestration outcomes. All
measurements are from the deployed prototype with real devices.

\begin{table}[t]
 \centering
 \caption{Orchestration evidence from the deployed prototype.
 End-to-end time includes LLM inference at each agent hop.}
 \label{tab:evidence}
 \begin{tabular*}{\columnwidth}{@{\extracolsep{\fill}}lr@{}}
 \toprule
 \textbf{Metric} & \textbf{Result} \\
 \midrule
 \emph{Scene 1: Intent-Driven Coordination} & \\
 Task completion rate & 4/5\textsuperscript{\dag} \\
 Median end-to-end latency & 8\,s \\[0.2em]
 \emph{Scene 2: Conflict Resolution} & \\
 Conflicts correctly detected / resolved & 5/5 \\[0.2em]
 \emph{Scene 3: Freshness and Authorization Verification} & \\
 Stale commands correctly rejected & 5/5 \\
 Expired / invalid leases rejected & 5/5 \\
 False rejections & 0 \\[0.2em]
 \emph{Cross-cutting} & \\
 Events persisted to Git & 153/153 \\
 Lease validation overhead (p95) & $<$0.01\,ms \\
 \bottomrule
 \\[-0.5em]
 \multicolumn{2}{@{}l@{}}{\footnotesize\textsuperscript{\dag}Vision-grounding error on mobile agent (misidentified app icon); protocol} \\
 \multicolumn{2}{@{}l@{}}{\footnotesize correctly logged for post-hoc diagnosis. All metrics are from live runs with LLM} \\
 \multicolumn{2}{@{}l@{}}{\footnotesize inference; the artifact replays fixed MQTT traces for deterministic reproducibility.} \\
 \end{tabular*}
\end{table}

End-to-end coordination time is dominated by LLM inference latency at
each agent hop (typically 1--5\,s per hop), not by the orchestration
infrastructure itself. Lease validation overhead (p95) was $<$0.01\,ms. This decomposition matters: orchestration overhead is
negligible, while freshness and lease checks are local metadata
comparisons. Inference latency remains the dominant external variable
and should decrease as edge-deployable models improve.

\paragraph{Testbed.}
Root agent: Mac mini (M4 ARM64, 16\,GB RAM).
Manager and librarian: Intel NUC11 (x64, 4 cores, 7.5\,GB RAM, Ubuntu
24.04). Mobile agent: Pixel 4 (Android 14). Network: Tailscale mesh
VPN. Broker: Mosquitto 2.0 on Mac mini. Devices: Philips Hue lights,
JBL speakers, Reolink camera, LG smart TV. LLM inference: Anthropic API
(Claude Opus 4.6), Google Vertex AI (Gemini 3 Pro).

\paragraph{Limitations.}
Measurements are from a single testbed with 4 agents and $\sim$10 devices. Scalability to larger deployments is untested. LLM inference uses hosted APIs; latency depends on provider load and is reported
transparently as end-to-end timing. We assume an honest-but-crashing failure model; compromised-agent and replay-attack threats are out of scope and complementary to runtime defenses such as~\cite{prompt-injection,agentdog,pasb}.

\paragraph{Artifact.}
The artifact bundles broker configuration, OpenClaw agent templates, the ground-truth Git repository, lease-validation logic, and replay scripts. Replay drives orchestration via fixed-timing MQTT traces (LLM inference excluded), verifying safety properties without hardware; device actuation requires the testbed.   

\section{Related Work}
\label{sec:related}

OpenAI's Agents SDK~\cite{openai-agents}, Microsoft's
AutoGen~\cite{ms-autogen}, LangChain's LangGraph~\cite{langchain-multiagent},
and Anthropic's agent teams~\cite{anthropic-agents} provide mature
orchestration primitives, but they largely leave cross-session state,
device-specific authority, and recovery semantics to the application.
In the smart-home domain, prior work studies conflict detection across
co-located IoT applications~\cite{smart-home-mas} and interoperability
across protocol stacks~\cite{iot-interop}. \textsc{HearthNet} builds on
these threads with a deployed edge architecture in which agents run on
separate machines, coordinate through MQTT, externalize state to Git,
and execute through root-issued actuation leases.

Indirect prompt injection~\cite{prompt-injection},
AgentDoG~\cite{agentdog}, OpenClaw Attack Benchmark~\cite{pasb}, and the CPS security survey
by Hatami et al.~\cite{sentinel} address unsafe model behavior and
runtime trust, while NIST IoT guidance~\cite{nist-iot} and provenance
frameworks such as SLSA and in-toto~\cite{slsa,intoto} address device
and supply-chain integrity. \textsc{HearthNet} is complementary: it
targets \emph{coordination safety}, requiring each state-changing
command to be both fresh with respect to shared state and authorized
with respect to an explicit policy snapshot before it can reach a
device.

\section{Conclusion}

 We presented \textsc{HearthNet}, an edge multi-agent orchestration system for smart homes that combines persistent role-specialized agents on OpenClaw~\cite{openclaw}, MQTT coordination, and Git-backed shared state under root-mediated freshness checks and policy-bound actuation leases. The deployed prototype across three classes of edge hardware demonstrates that practical persistent multi-agent coordination on commodity devices is achievable today.

 Future work replaces hosted LLM APIs with locally deployed models to eliminate the cloud dependency and integrates SENTINEL-style monitoring agents~\cite{gosmar2025sentinel} for stronger runtime trust.

\bibliographystyle{ACM-Reference-Format}
\bibliography{references}

@misc{anthropic-agents,
  author = {{Anthropic}},
  title = {Effective Harness for Long Running Agents},
  year = {2025},
  howpublished = {\url{https://www.anthropic.com/engineering/effective-harnesses-for-long-running-agents}},
  note = {Accessed 2026-03-11}
}

@misc{anthropic-long-running,
  author = {{Anthropic}},
  title = {Claude Code: Best Practices},
  year = {2026},
  howpublished = {\url{https://code.claude.com/docs/en/best-practices}},
  note = {Accessed 2026-03-11}
}

@misc{ms-autogen,
  author = {{Microsoft}},
  title = {AutoGen: A framework for building AI agents and applications},
  year = {2025},
  howpublished = {\url{https://microsoft.github.io/autogen/dev/index.html}},
  note = {Accessed 2026-03-11}
}

@misc{ms-agentframework-migration,
    author = {{Microsoft}},
    title  = {{AutoGen} to {Microsoft} Agent Framework Migration Guide},
    year ={2026},
    howpublished = {\url{https://learn.microsoft.com/en-us/agent-framework/migration-guide/from-autogen/}},
    note = {Accessed 2026-03-11}            
  }

@misc{langchain-multiagent,
  author = {{LangChain}},
  title = {LangChain: Multi-agent},
  year = {2025},
  howpublished = {\url{https://docs.langchain.com/oss/python/langchain/multi-agent}},
  note = {Accessed 2026-03-11}
}

@misc{openai-agents,
  author = {{OpenAI}},
  title = {Agents SDK},
  year = {2025},
  howpublished = {\url{https://developers.openai.com/api/docs/guides/agents-sdk/}},
  note = {Accessed 2026-03-11}
}

@book{nist-iot,
  title={Foundational cybersecurity activities for IoT device manufacturers},
  author={Fagan, Michael and Megas, Katerina N and Scarfone, Karen and Smith, Matthew},
  year={2020},
  publisher={US Department of Commerce, National Institute of Standards and Technology}
}

@article{agentdog,
  title={AgentDoG: A Diagnostic Guardrail Framework for AI Agent Safety and Security},
  author={Liu, Dongrui and Ren, Qihan and Qian, Chen and Shao, Shuai and Xie, Yuejin and Li, Yu and Yang, Zhonghao and Luo, Haoyu and Wang, Peng and Liu, Qingyu and others},
  journal={arXiv preprint arXiv:2601.18491},
  year={2026}
}

@article{sentinel,
  title={Securing AI Agents in Cyber-Physical Systems: A Survey of Environmental Interactions, Deepfake Threats, and Defenses},
  author={Hatami, Mohsen and Pham, Van Tuan and Lakadawala, Hozefa and Chen, Yu},
  journal={arXiv preprint arXiv:2601.20184},
  year={2026}
}

@article{pasb,
  title={From Assistant to Double Agent: Formalizing and Benchmarking Attacks on OpenClaw for Personalized Local AI Agent},
  author={Wang, Yuhang and Xu, Feiming and Lin, Zheng and He, Guangyu and Huang, Yuzhe and Gao, Haichang and Niu, Zhenxing and Lian, Shiguo and Liu, Zhaoxiang},
  journal={arXiv preprint arXiv:2602.08412},
  year={2026}
}

@inproceedings{prompt-injection,
  title={Not what you've signed up for: Compromising real-world llm-integrated applications with indirect prompt injection},
  author={Greshake, Kai and Abdelnabi, Sahar and Mishra, Shailesh and Endres, Christoph and Holz, Thorsten and Fritz, Mario},
  booktitle={Proceedings of the 16th ACM workshop on artificial intelligence and security},
  pages={79--90},
  year={2023}
}

@misc{slsa,
  title={Supply-chain Levels for Software Artifacts},
  author={SLSA Authors},
  year={2024},
  howpublished={\url{https://slsa.dev/}},
  note={Accessed 2026-03-11} 
}

@inproceedings{intoto,
  title={in-toto: Providing farm-to-table guarantees for bits and bytes},
  author={Torres-Arias, Santiago and Afzali, Hammad and Kuppusamy, Trishank Karthik and Curtmola, Reza and Cappos, Justin},
  booktitle={28th USENIX Security Symposium (USENIX Security 19)},
  pages={1393--1410},
  year={2019}
}

@misc{openclaw,
  author = {Steinberger, Peter and {OpenClaw Community}},
  title = {OpenClaw: Personal AI Assistant},
  year = {2025},
  howpublished = {\url{https://openclaw.ai/}},
  note = {Accessed 2026-03-11}
}

@misc{home-assistant,
  author = {{Home Assistant Contributors}},
  title = {Home Assistant: Open-Source Home Automation},
  year = {2024},
  howpublished = {\url{https://www.home-assistant.io/}},
  note = {Accessed 2026-03-11}
}

@misc{aws-device-shadow,
  author = {{Amazon Web Services}},
  title = {AWS IoT Device Shadow service},
  year = {2024},
  howpublished = {\url{https://docs.aws.amazon.com/iot/latest/developerguide/iot-device-shadows.html}},
  note = {Accessed 2026-03-11}
}

@article{smart-home-mas,
  title={A survey on conflict detection in IoT-based smart homes},
  author={Huang, Bing and Chaki, Dipankar and Bouguettaya, Athman and Lam, Kwok-Yan},
  journal={ACM Computing Surveys},
  volume={56},
  number={5},
  pages={1--40},
  year={2023},
  publisher={ACM New York, NY}
}

@article{iot-interop,
  title={Internet of things: A survey on enabling technologies, protocols, and applications},
  author={Al-Fuqaha, Ala and Guizani, Mohsen and Mohammadi, Mehdi and Aledhari, Mohammed and Ayyash, Moussa},
  journal={IEEE communications surveys \& tutorials},
  volume={17},
  number={4},
  pages={2347--2376},
  year={2015},
  publisher={Ieee}
}

@article{gosmar2025sentinel,
  title={Sentinel agents for secure and trustworthy agentic ai in multi-agent systems},
  author={Gosmar, Diego and Dahl, Deborah A},
  journal={arXiv preprint arXiv:2509.14956},
  year={2025}
}

\end{document}